\documentstyle[11pt,aaspp4]{article}
\input epsf.tex
\newcommand{\onu}{\Omega_\nu}
\newcommand{\oc}{\Omega_c}
\newcommand{\om}{\Omega_{\rm m}}
\newcommand{\ob}{\Omega_{\rm b}}
\newcommand{\ov}{\Omega_\Lambda}
\newcommand{\qrms}{Q_{\rm rms-PS}}

\newcommand{\xinl}{\bar\xi_{\rm nl}}
\newcommand{\xil}{\bar\xi_{\rm l}}
\newcommand{\rnl}{r_{\rm nl}}
\newcommand{\rl}{r_{\rm l}}
\newcommand{\knl}{k_{\rm nl}}
\newcommand{\kl}{k_{\rm l}}
\newcommand{\Pnl}{P_{\rm nl}}
\newcommand{\Pl}{P_{\rm l}}
\newcommand{\Dnl}{\Delta_{\rm nl}}
\newcommand{\Dl}{\Delta_{\rm l}}

\begin{document}
\title{Nonlinear Cosmological Structure Formation}
\author{Chung--Pei Ma}
\affil{Department of Physics and Astronomy, University of Pennsylvania, 
Philadelphia, PA 19104; cpma@strad.physics.upenn.edu}

\begin{abstract}
I present a general discussion of the evolution and model-dependence
of both the linear and nonlinear power spectrum of density
fluctuations.  The features of the linear power spectrum in
cosmological models with cold dark matter (CDM) and cold+hot dark
matter (C+HDM) are reviewed, and useful analytical approximations are
summarized.  Cosmological numerical simulation results are then used
to illustrate the process of nonlinear gravitational clustering and to
compute the nonlinear power spectrum.  A new analytical approximation
relating the linear and nonlinear power spectrum in C+HDM models is
presented.
\end{abstract}

\section{Introduction}
The power spectrum $P(k)$ of matter fluctuations is a fundamental
quantity in cosmology.  It provides the most basic statistical measure
of gravitational clustering, and an accurate determination of the
power spectrum is among the most important goals of every major galaxy
survey.  Since each cosmological model has its distinct power
spectrum, one can hope to obtain crucial information such as the matter
content of the Universe and the nature of the primordial fluctuations
by comparing the observed power spectrum with theoretical predictions
for various plausible models.

Measurements of the power spectrum require extensive galaxy surveys
covering a large volume of space.  It is only recently that such
surveys have been completed and are starting to reveal intriguing
results.  For instance, much interest has been generated by the
compilation of the power spectrum determined from the APM and IRAS
galaxy catalogs and the effort to reconstruct the underlying linear
mass power spectrum (Peacock \& Dodds 1994; Peacock 1997).  The
recently completed Las Campanas survey of over 23,000 galaxies has
yielded a power spectrum incompatible with the standard cold dark
matter (CDM) model but still consistent with a class of open CDM,
cold+hot dark matter (C+HDM), and tilted CDM models (Lin et al. 1996).
Larger ongoing projects such as the CfA Century, the 2dF, and the
Sloan surveys promise to provide measurements with reduced error bars
and data points on larger scales for better constraints.

On the theoretical side, calculations of the power spectrum falls into
two regimes: the linear and the nonlinear, which are characterized by
the amplitude of the density fluctuations $\delta=\delta\rho/\rho \ll
1$ and $>1$, respectively.  Since the observed power spectrum spans
both linear and nonlinear regimes, theoretical calculations in both
regimes must be performed before all data points can be fully utilized
as constraints.  In the linear regime, the computation of the power
spectrum has become standard practice.  It is obtained by integrating
the coupled, linearized Einstein and Boltzmann equations that describe
the evolution of the metric and density perturbations throughout the
history of the Universe.  Several groups have written numerical codes
for such calculations, and a publicly-available version is described
in Ma \& Bertschinger (1995) and placed at http://arcturus.mit.edu.
This code computes, in both synchronous and conformal Newtonian
gauges, the evolution of the phase-space distributions of photons,
baryons, cold dark matter, and both massless and massive
neutrinos.  It is therefore applicable to most CDM, C+HDM, and CDM
with a cosmological constant (LCDM) models.  To gain deeper insight
and for convenience, analytical approximations for the linear
power spectra in these models have also been published.  Details
about the linear power spectrum are discussed in Section 2.

Gravitational clustering, however, eventually becomes a nonlinear
process.  The existence of galaxies and clusters is a manifestation of
the nonlinear nature of gravity.  Determining the nonlinear power
spectrum $\Pnl$ is therefore an important task.  Not surprisingly,
$\Pnl$ is more difficult to obtain than the linear power spectrum.
Higher-order perturbation theories help to extend the range of
validity to the quasi-linear regime, but the fully nonlinear power
spectrum can be calculated only from numerical simulations.  Thus far,
theoretical predictions of $\Pnl$ have been carried out for only a few
models at limited epochs because it is difficult and time-consuming to
perform numerical simulations with sufficient dynamic range to allow
calculations of $\Pnl$ over a wide range of scales.
Section 3 summarizes some recent work and presents new 
analytic and numerical results for nonlinear clustering in C+HDM models.

%\section{Neutrino Phase Space}
\section{Linear Power Spectrum}
The power spectrum $P(k)$ of matter fluctuations 
is related to the density field $\delta$ in $k$-space by
\begin{equation}
	\left< \delta(\vec{k}) \delta^* (\vec{k}^\prime) \right>
	= P(k) \delta_D(\vec{k}-\vec{k}^\prime)\,, 
\end{equation} 
where $\delta_D$ is the Dirac-delta function. The Fourier transform of
the power spectrum is the two-point correlation function
$\xi(r)=\left<\delta(\vec{x_1}) \delta(\vec{x_2})  \right>$, where
$r\equiv |\vec{x_1}-\vec{x_2}|$. For a Gaussian density field (as
predicted in most inflationary theories), its statistical property is
entirely determined by $P(k)$.

The shape of the linear power spectrum depends on the cosmological
parameters assumed in a given theory of structure formation.  It is
governed by the linear perturbation theory of gravitational
clustering, and can be computed by integrating the coupled, linearized
Einstein, Boltzmann, and fluid equations for the metric and density
perturbations.  Some important parameters that affect the shape of the
power spectrum are: (1) {\it Primordial spectral index $n$}:
$P(k)\propto k^n$.  An example is the Harrison-Zeldovich (Harrison
1970; Zeldovich 1972) spectrum which takes $n=1$.  Most versions of
inflationary models also predict nearly $n=1$ power spectra. (2) {\it
Matter-radiation equality time}: The equality time $t_{\rm eq}$ is
defined to be the epoch in the thermal history of the Universe when
the energy density in matter equals that in radiation.  The equality
redshift scales as $1+z_{\rm eq} \propto \om h^2$, where $\om$ is the
density parameter in matter and $h$ is the Hubble constant in units of
100 km/s/Mpc.  This parameter controls the location of the peak of
$P(k)$ in CDM-type models since the density perturbations with
wavenumbers $k > k_{\rm eq}$ enter the horizon in the
radiation-dominated era and cannot grow appreciably, whereas
perturbations with $k < k_{\rm eq}$ enter the horizon in the
matter-dominated era and can grow unimpeded.  (3) {\it Nature of dark
matter}.  The pure CDM model, for example, exhibits a characteristic
$P(k)\propto k^{-3}$ slope at high $k$, while the power spectrum for
the pure hot dark matter (HDM) model is cut off exponentially below
the free-streaming scale of the massive neutrinos due to the phase
mixing in the neutrino phase-space distribution.  The hybrid C+HDM
models, which are parameterized by the neutrino mass density $\onu$
and CDM mass density $\Omega_c=1-\onu-\Omega_b$ (where $\Omega_b$ is
the mass density in baryons), exhibit intermediate behavior.  The
suppression in the power in both the cold and hot components due to
neutrino free-streaming generally increases with increasing $\onu$.

The solid curves in Figure 1 show the linear power spectra for the pure
CDM and three C+HDM models with $\onu=0.1, 0.2$, and 0.3.  The
corresponding neutrino masses in the four models are 0, 2.3, 4.6, and
7 eV, respectively.  (Only one of the three species of neutrinos is
assumed massive.)  The models have a total matter density of
$\Omega_m=1$ and $h=0.5$, and all are normalized to the COBE rms
quadrupole $\qrms=18\,\mu K$ (Gorski et al. 1996), as evidenced by the
convergence of the curves at small $k$.  The differences at large $k$
reflect the distinct properties of cold and hot dark matter: the
larger $\onu$ is, the weaker is the gravitational clustering on small
length scales.  For comparison, a COBE-normalized low-density model
with $\Omega_m=0.3$, a cosmological constant
$\Omega_\Lambda=0.7$, and $h=0.75$ is also shown (dashed curve).  

Analytical approximations with less than $10$\% error are very
convenient as an input for many calculations.  The pure CDM model is
well-approximated by
\begin{eqnarray}
        P_c(q,a,\onu=0) &=& A\,a^2\,k^n \nonumber\\
        &\times &\left[{\ln(1+\alpha_1 q)\over \alpha_1 q}\right]^2 
         {1\over [1+\alpha_2 q+(\alpha_3 q)^2+(\alpha_4 q)^3
        +(\alpha_5\,q)^4]^{1/2}}\,,
        \quad q={ k\over \Gamma h} \,, 
\label{bbks}
\end{eqnarray} 
where $A$ is the overall normalization factor, $a$ is the expansion
factor, $n$ is the primordial spectral index, $k$ is the wavenumber in
units of Mpc$^{-1}$, and $\alpha_1=2.34,\alpha_2=3.89,\alpha_3=16.1,
\alpha_4=5.46$, and $\alpha_5=6.71$ (Bardeen et al.~1986).  The
normalization $A$ can be determined from the 4-year data from the COBE
satellite, and $A=2420\,h^{-4}$ Mpc$^4$ for $n=1$ flat models (Bunn \&
White 1997).  The shape parameter $\Gamma$ characterizes the
dependence on cosmological parameters, and a good approximation is
found to be given by $\Gamma=\om h/\exp[\ob(1+1/\om)]$ (Efstathiou et
al. 1992; Sugiyama 1995).
The error in the approximation for the standard CDM model with
$\ob=0.05$, for example, is $< 10$\%.  A higher-accuracy fit for this
much-studied model can be achieved by setting $\ob=0$ in
equation~(\ref{bbks}) (i.e.  setting $\Gamma=\om h$) and modifying the
coefficients to $\alpha_1=2.205, \alpha_2=4.05, \alpha_3=18.3,
\alpha_4=8.725$, and $\alpha_5=8.0$.  The fractional error relative to
the direct numerical result is then reduced to smaller than 1\% for
$k<40\,h$ Mpc$^{-1}$ (Ma 1996).

The linear power spectra for the C+HDM models require additional
treatment since the effect of neutrino free-streaming on the shape of
the power spectrum is both time- and scale-dependent.  It is found (Ma
1996) that by introducing a second shape parameter
\begin{equation}
        \Gamma_\nu=a^{1/2}\onu h^2
\end{equation}
to characterize the neutrino free-streaming distance,
accurate approximations to the linear power spectra
in C+HDM models can be obtained.   The cold and hot components
clearly have different power spectra due to their different thermal
properties.  For the CDM component in C+HDM models,
a good approximation is given by
\begin{equation}
    P_c(k,a,\onu) = P_c(k,a,\onu=0)
        \left( { 1+b_1\,x^{b_4/2}+b_2\,x^{b_4} \over 1+b_3\,x_0^{b_4} }
        \right)^{\onu^{1.05}}\,,
   \quad x = {k\over \Gamma_\nu} \,, \quad x_0=x(a=1)\,,
\label{pc}
\end{equation} 
where $P_c(k,a,\onu=0)$ for the pure CDM model is given by
equation~(\ref{bbks}), and $b_1=0.01647, b_2=2.803\times 10^{-5},
b_3=10.90$, and $b_4=3.259$ for $k$ in units of Mpc$^{-1}$.  The
functional form for the ratio $P_c(k,a,\onu)/P_c(k,a,0)$ is chosen to
have the asymptotic behavior $\propto a^{2(f_\infty-1)}$, which can be
derived analytically, and $f_\infty= (5\sqrt{1-24\onu/25}-1)/4$
is the asymptotic growth rate for $k\rightarrow \infty$, where
$1-f_\infty \propto \onu^{1.05}$ is a good approximation.  For the HDM
component in C+HDM models, an accurate approximation is given by
\begin{equation}
    P_\nu(k,a,\onu) = P_c(k,a,\onu)
        \left( {e^{-c_1\,x} \over 1 + c_2\,x^{1/2} + c_3\,x
	+ c_4\,x^{3/2} + c_5\,x^2 } \right) \,,
   \quad x = {k\over \Gamma_\nu\,h} \,,
\label{pnu}
\end{equation} 
where $c_1=0.0015, c_2=-0.1207, c_3=0.1015, c_4=-0.01618$, and
$c_5=0.001711$ for $k$ in units of Mpc$^{-1}$.

It is often useful to have an analytic approximation for the
density-weighted power spectrum $P(k)=\{\onu P_\nu^{1/2} + (1-\onu)
P_c^{1/2}\}^2$ that measures the total gravitational fluctuations
contributed by the two components.  Here the CDM and baryons have been
assumed to have the same power (i.e., $P_c=P_b$), which is a good
approximation for the range of redshifts and $\Omega_b$ studied in
this paper.  The functional form used for the CDM spectrum $P_c$ in
equation~(\ref{pc}) works well here, and a good approximation 
for the density-weighted power spectrum in C+HDM models is given by
\begin{equation}
   {P(k,a,\onu)\over P_c(k,a,\onu=0)}=g(x,\onu)=
        \left( { 1+d_1\,x^{d_4/2}+d_2\,x^{d_4} \over 1+d_3\,x_0^{d_4} }
        \right)^{\onu^{1.05}}\,,
   \quad x = {k\over \Gamma_\nu} \,, \quad x_0=x(a=1)\,,
\label{pave}
\end{equation} 
where $P_c(k,a,\onu=0)$ again is given by equation~(2), and
the coefficients are $d_1=0.004321, d_2=2.217\times 10^{-6},
d_3=11.63$, and $d_4=3.317$ for $k$ in units of Mpc$^{-1}$.

It should be noted that the C+HDM power spectra do not obey the simple
evolution $P(k)\propto a^2$ in equation~(2) for the flat pure CDM
model.  The free-streaming of massive neutrinos slows down with time,
allowing the neutrinos to cluster gravitationally and the neutrino
density perturbations to grow on increasingly smaller scales.  As a
result, the growth of the C+HDM power spectra is both scale- and
time-dependent.  This effect is taken care of by the time-dependent
parameter $\Gamma_\nu$ in equation~(3), which is built in in
equations~(4)-(6) via the scaled variable $x$.

\section{Nonlinear Gravitational Clustering}
\subsection{Numerical Simulations}
The linear theory of gravitational clustering is an elegant and
powerful theory.  It describes accurately the growth of density
perturbations from the early Universe until a redshift of $\sim 100$,
and the solution to the theory involves straightforward
time-integration of a set (albeit a large set) of coupled ordinary
differential equations.  Gravitational clustering, however, eventually
becomes a nonlinear process, and the study of the fully nonlinear
process ultimately relies on numerical simulations.

Cosmological simulations can be roughly divided into two categories:
(1) $N$-body simulations that deal only with dissipationless
gravitational interactions of dark matter; and (2) hydrodynamical
simulations that model gaseous dissipation via cooling and heating
processes in addition to the gravitational interactions among the dark
matter and gas.  Depending on the models, the simulations are started
at a redshift between 20 and 100 when the rms density fluctuations are
well below unity, and the initial conditions of the simulations are
generated from the linear power spectrum at that epoch.

As an illustration of gravitational clustering in collisionless
$N$-body simulations, Figure~2 shows the projection of the smoothed
matter distribution at four redshifts, $z=5, 3, 1.5$, and 0, from an
$N$-body simulation of the $\onu=0.1$ C+HDM model.  The model assumes
$\oc=0.85$, $\ob=0.05$, and $h=0.5$, and is normalized to the COBE
quadrupole $\qrms=18\,\mu K$ (Gorski et al. 1996).  The simulation box
is 100 Mpc comoving on a side, and the force softening is 50 kpc.  A
total of $128^3$ cold and $128^3$ hot particles are used.  Each panel
in Figure~2 is 100$\times$100 Mpc comoving, and shows the projection
along one axis of the entire simulation box.  Periodic boundary
conditions are adopted in the simulation, so for example, the dense
clumps near the central left and right edges of the box in the $z=0$ panel
(lower right) are part of the same cluster of dark matter halos.  The
darkest several halos at $z=0$ all have masses above $10^{14}
M_\odot$.

The power spectrum of the density fluctuations offers the lowest-order
measurement of the gravitational clustering exhibited in Figure~2.  In
Figure~3, we plot the corresponding nonlinear power spectra computed
from the particle spatial distributions at $z=3, 1.5$, and 0 shown in
Figure~2.  The linear power spectra given by equation~(6) are also
shown for comparison.  The hierarchical, or ``bottom-up'', nature of
gravitational collapse in these models is evident: the high-$k$ modes
(i.e. small length scales) have become strongly nonlinear, while the
low-$k$ modes are still following the linear power spectrum.  The fact
that the three lowest $k$ modes are still linear at $z=0$ ensures that
our choice of the simulation box size (100 Mpc) is large enough to
include all waves that have gone nonlinear at the present epoch.  It
is also interesting to note that the point of departure from linearity
moves to the left of the figure as $z$ decreases, indicating that
objects become nonlinear on increasingly larger length scales as the
Universe evolves.

It is also instructive to compare the nonlinear structures at the same
epoch in different cosmological models.  Figure~4 shows the smoothed
matter distribution at $z=1.5$ from simulations of the four models in
Figure~1: a LCDM, and C+HDM with $\onu=0.1, 0.2$, and 0.3 (clockwise
starting from upper left).  The LCDM model has $\om=0.3$, $\ov=0.7$,
and $h=0.75$, and the three C+HDM models have $h=0.5$, $\ob=0.05$, and
$\oc+\onu+\ob=1$.  All models are COBE-normalized, and all simulations
are performed with the same box size and force resolution as the
$\onu=0.1$ run described earlier.  (The only exception is that the
$\onu=0.2$ and 0.3 simulations used $10\times 128^3$ instead of
$128^3$ particles to represent the HDM component.)  The initial
conditions for all four simulations are generated with the same random
phases; structures therefore appear at similar locations in all
panels.  The effect of $\onu$ on structure formation, however, is
striking: the higher $\onu$ is, the fewer collapsed objects there are
at a given epoch.  This trend simply reflects the decrease in the
small-scale linear power shown in Figure~1, which results from
neutrino free-streaming.  Note that the model with $\onu=0.3$ (lower
left panel), which corresponds to a neutrino mass of 7 eV, has very
few structures even at $z=1.5$.  It is also interesting to note that
the upper two panels of Figure~4 for the LCDM and $\onu=0.1$ C+HDM
models look very similar.  This is because the linear power spectra
for the two models are in fact very similar at $k >0.1\,h$ Mpc$^{-1}$,
as shown in Figure~1.  The corresponding linear and nonlinear power
spectra at $z=1.5$ for the three C+HDM models in Figure~4 are shown in
Figure~5.

The power spectra shown thus far for the C+HDM models are the
density-averaged $P(k)=\{\onu\,P_\nu^{1/2}+(1-\onu)\,P_c^{1/2}\}^2$,
where $P_\nu$ and $P_c$ are the individual power spectra for the hot
and cold components.  (It has been assumed $P_c= P_{\rm baryon}$,
which is a good approximation for the redshift range of interest
here.)  The two components evolve distinctly due to their different
thermal velocities, and the details of the shape and the growth of the
linear power spectra are discussed in Ma (1996).  Here, the nonlinear
power spectra for the separate components are presented in Figure~6.
For clarity, only two epochs, $z=3$ and 0, are plotted.  Notice how at
$z=3$, the HDM spectrum remains linear to almost $k=2\,h\,$Mpc$^{-1}$
but the CDM remains linear to only $k\sim 0.5\,h$Mpc$^{-1}$, while at
$z=0$, both components become nonlinear at $k\approx
0.3\,h\,$Mpc$^{-1}$.  This is largely due to the slowing down of the
neutrinos, which makes it easier for the neutrino particles to fall
into the CDM potential wells at later times, and therefore become
nonlinear at smaller $k$.  More precisely, the median thermal velocity
of neutrinos is
\begin{equation}
	v\approx 3 (1+z) {k_B T_{0,\nu}\over m_\nu c }
	= 15 (1+z){10\,{\rm eV}\over m_\nu}\,{\rm km/s} \,,
\end{equation}
so the neutrinos in the $\onu=0.1$ model ($m_\nu=2.3$ eV) have slowed down 
by a factor of 4 since $z=3$ to about 60 km/s today.

\subsection{Analytical Approximations}

Just as it was useful to cast the linear power spectra in simple
functional forms for a wide range of models (see equations~(1)-(6)),
analytical approximations to the nonlinear power spectrum can also
provide valuable insight to the process of gravitational clustering.
It is, however, more difficult to compute the nonlinear $\Pnl$ because
the numerical simulations that have sufficient dynamic range to allow
calculations of $\Pnl$ over a wide range of scales are generally much
more CPU-intensive than the integration of the linearized Boltzmann
equations.  The behavior of the power spectrum in the nonlinear regime
therefore is less well-understood.

An early attempt to relate linear and nonlinear quantities was focused
on the spatially-averaged two-point correlation function
$\bar{\xi}(r)$ in models with $\om=1$ and a power-law power spectrum
(Hamilton et al. 1991).  It was found that if a given nonlinear scale
$\rnl$ is identified with its pre-collapsed linear scale $\rl$ by
$\rl^3=\rnl^3(1+\xinl)$, then there exists a simple, universal
function relating the linear and nonlinear two-point correlation function:
\begin{equation}
 \xinl(\rnl) = F[\xil(\rl)]\,.
\end{equation}
At that time, this transformation appeared to be magically independent
of the spectral index $n$ of the linear power spectrum assumed in the
model, and the generality of this formula rendered the task of
reconstructing the primordial spectrum from the observed nonlinear
clustering of galaxies less complicated.  However, further tests
against numerical simulations showed that this formula worked well
only for spectral index $n>-1$, and it erred by factors up to 3 and 10
for $n=-1.5$ and $-2$, respectively (Jain et al. 1995).  In
particular, the nonlinear $\xinl$ was found to rise more sharply with
increasing $\xil$ for models with more negative $n$, so Jain et
al. introduced an $n$-dependent formula to accommodate this feature.
For the more realistic cosmological models such as the CDM where the
spectral index is a function of scale, they proposed to use
an effective index $n_{\rm eff}$, defined to be $n_{\rm eff}=d\ln
P/d\ln k_0$, where $k_0$ is the scale at which the rms mass
fluctuation $\sigma(R=1/k_0)$ is unity.  The index $n_{\rm eff}$
therefore represents the slope of the power spectrum at a scale where
nonlinear effects are becoming important.  Later work extended the
analysis to models with varying matter density $\om$ and cosmological
constant $\ov$ (Peacock \& Dodds 1996), where instead of using an
effective index for all scales, the local slope $n(k)=d\ln P/d\ln k$
was adopted.

Here we examine the nonlinear mapping of the power spectrum in 
the C+HDM models, which has not been explored yet.
Figure~7 shows how the nonlinear density variance
$\Dnl=4\pi\knl^3\,\Pnl(\knl)$ diverges from the linear
$\Dl=4\pi\kl^3\,\Pl(\kl)$, where the linear and nonlinear wavenumbers
are related by $\kl^3=\knl^3/(1+\Dnl)$.  The squares are obtained from
simulations of the $\onu=0.2$ (left curve) and 0.1 (right) C+HDM
models.  The curves generally obey the asymptotic condition $\Dnl=\Dl$
for $\Dl \ll 1$, and $\Dnl\propto \Dl^{1.5}$ in the highly-nonlinear,
stable clustering regime.  The mapping, however, is clearly {\it not}
independent of cosmology: the larger $\onu$ is in a model, the faster
$\Dnl/\Dl$ increases at $\Dl\sim 1$.  This trend reflects the
$n$-dependence pointed out by Jain et al. (1995) and can be explained
by the different shapes of $\Pl$ shown in Figure~1, where the models
with higher $\onu$ have less power and hence more negative slope at
high $k$.  Since all models are normalized to COBE and have the same
amplitude at low $k$, the critical $k_0$ (where $\sigma(k_0)=1$)
increases for larger $\onu$.  The effective index $n_{\rm eff}$ defined
at $k_0$ is therefore more negative for larger $\onu$, resulting in
the steeper rise of $\Dnl$ in Figure~7.

However, neither formula proposed by Jain et al. (1995) or Peacock \&
Dodds (1996) can be extended to the C+HDM models.  The dotted curves
in Figure~7 illustrate the large discrepancies in the Peacock-Dodds
fitting function when it is applied to the $\onu=0.1$ and 0.2 models.
In retrospect, it is not surprising their formulas do not apply:
Although the power spectrum at a given epoch is determined entirely
from the spatial distribution of the particles, the evolution of the
power spectrum depends on the particle velocities as well.  
A general formula for the linear to
nonlinear transformation therefore must depend on both the shape and
the growth rate of $P(k)$.  Since both formulas are designed for
models without massive neutrinos and depend only on the shape of
$P(k)$, they cannot be applied to C+HDM models.

Here we propose a new analytical approximation for the mapping of
linear and nonlinear power spectrum in CDM as well as C+HDM models:
\footnote{An improved approximation with a higher accuracy has since 
appeared in Ma (1998).}

\begin{equation}
        \Dnl(\knl)=\Delta_{\rm l}(\kl) 
        G\left[{\Delta_l}(\kl) \right] \,,
\end{equation}
where
\begin{equation}
       G(x)={1 + a_1 x^4 + a_2 x^8/g^{2.5} \over 
	1 + a_3 x^4 + a_4 x^{7.5}/g^2} \,,
\end{equation}
and the coefficients are $a_1=4756, a_2=384.6, a_3=3732$, and
$a_4=24.20$.  The functional form of $G$ is chosen to give the
appropriate asymptotic behavior $\Dnl \rightarrow \Dl$ in the linear
regime ($x\ll 1$) and $\Dnl \propto \Dl^{3/2}$ in the stable
clustering regime ($x\gg 1$).  The dependence of $G$ on $\onu$ comes
from the function $g$ of equation~(6), which gives the relative
amplitudes of the power spectra in the C+HDM models and the pure CDM
model.  This function is analogous to the commonly-used growth factor
$g(a=1,\om,\ov) = {5\over 2}\om
[\om^{4/7}-\ov+(1+\om/2)(1+\ov/70)]^{-1}$ for LCDM models (Lahav et
al. 1991; Carroll, Press \& Turner 1992).  They differ, however, in
that the growth factor is scale-independent in LCDM models but is a
function of scale in C+HDM models since neutrinos only affect the
growth below the free-streaming scale, as discussed in Section 2.

\section{Summary}
The power spectrum is a fundamental measure of gravitational
clustering in cosmology.  We have discussed the features and evolution
of both the linear and nonlinear power spectra in various cosmological
models.  

The linear power spectrum $P_l$ is calculated from time integration of
the coupled, linearized Einstein, Boltzmann, and fluid equations for
the metric and density perturbations.  The shape and growth of the
power spectrum depend on cosmological parameters such as the total
matter density $\om$, the neutrino fraction $\onu$, and the Hubble
constant $h$.  Simple analytical functions were presented that can
approximate the linear $P_l$ in both CDM and C+HDM models with
less than 10\% error.

The fully nonlinear power spectrum $P_{\rm nl}$ has to be computed
from numerical simulations.  Simulation output for CDM, LCDM, and
C+HDM models was presented to illustrate the sensitive dependence of
structure formation and evolution on cosmological parameters.  We also
discussed the present understanding of the relation between the linear
and nonlinear power spectra, and proposed a new approximation which
applies to CDM as well as C+HDM models.

The supercomputing time for this work was generously
provided by the National Scalable Cluster Project at the University of
Pennsylvania and the National Center for Supercomputing Applications.

\newpage
%\section*{Figure Captions}
%%%% Figure 1 %%%%
\begin{figure}
\epsfxsize=6.5truein 
\epsfbox{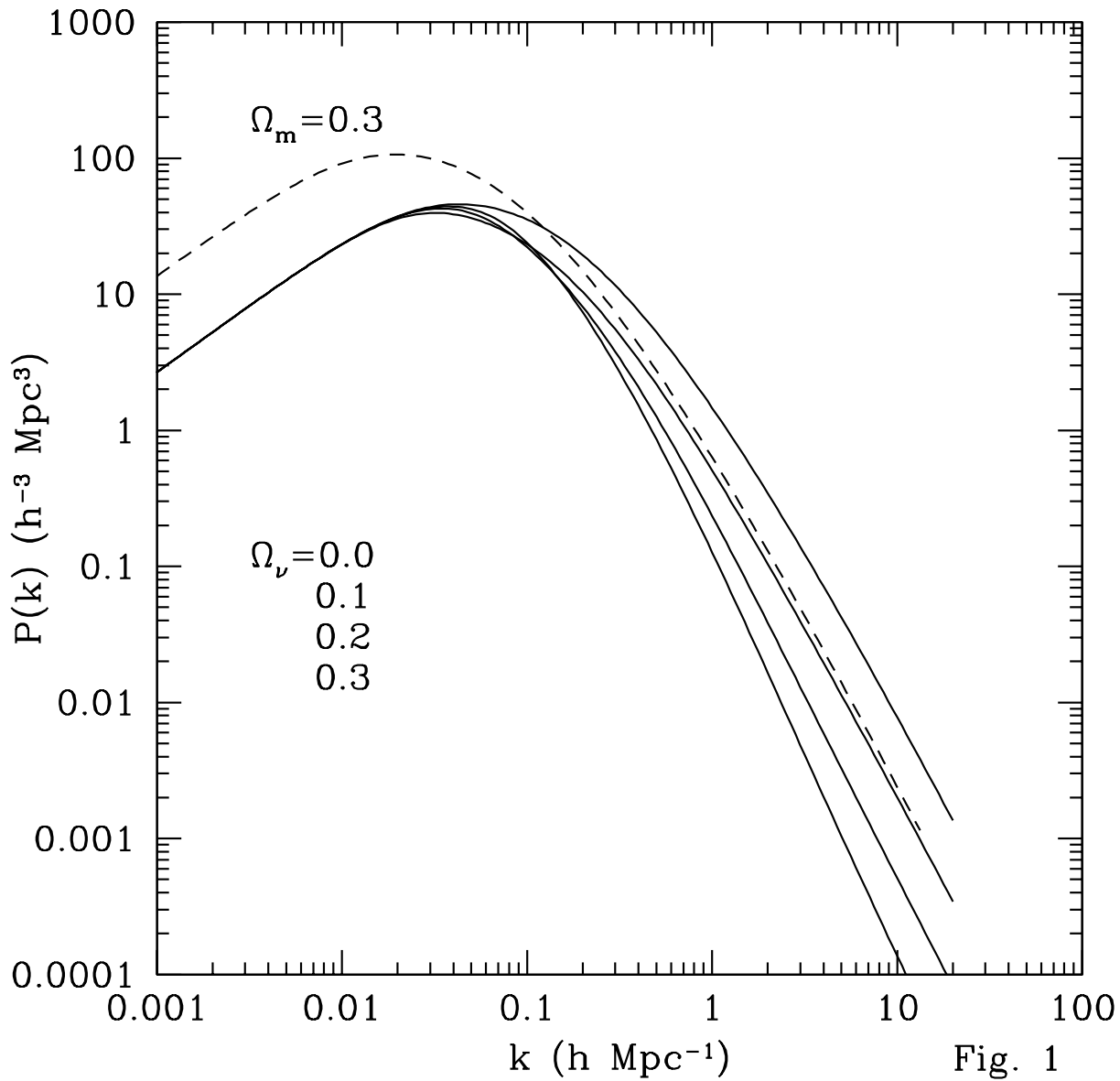}
\caption{The linear power spectrum of density fluctuations at $z=0$
for the standard CDM model (top solid), the C+HDM models with
$\onu=0.1, 0.2$, and 0.3 (lower solid), and the LCDM model with
$\Omega_m=0.3$ and $\Omega_\Lambda=0.7$ (dashed).  All are
COBE-normalized. }
\end{figure}

%%%% Figure 2 %%%%
\begin{figure}
\epsfxsize=6.5truein 
\epsfbox{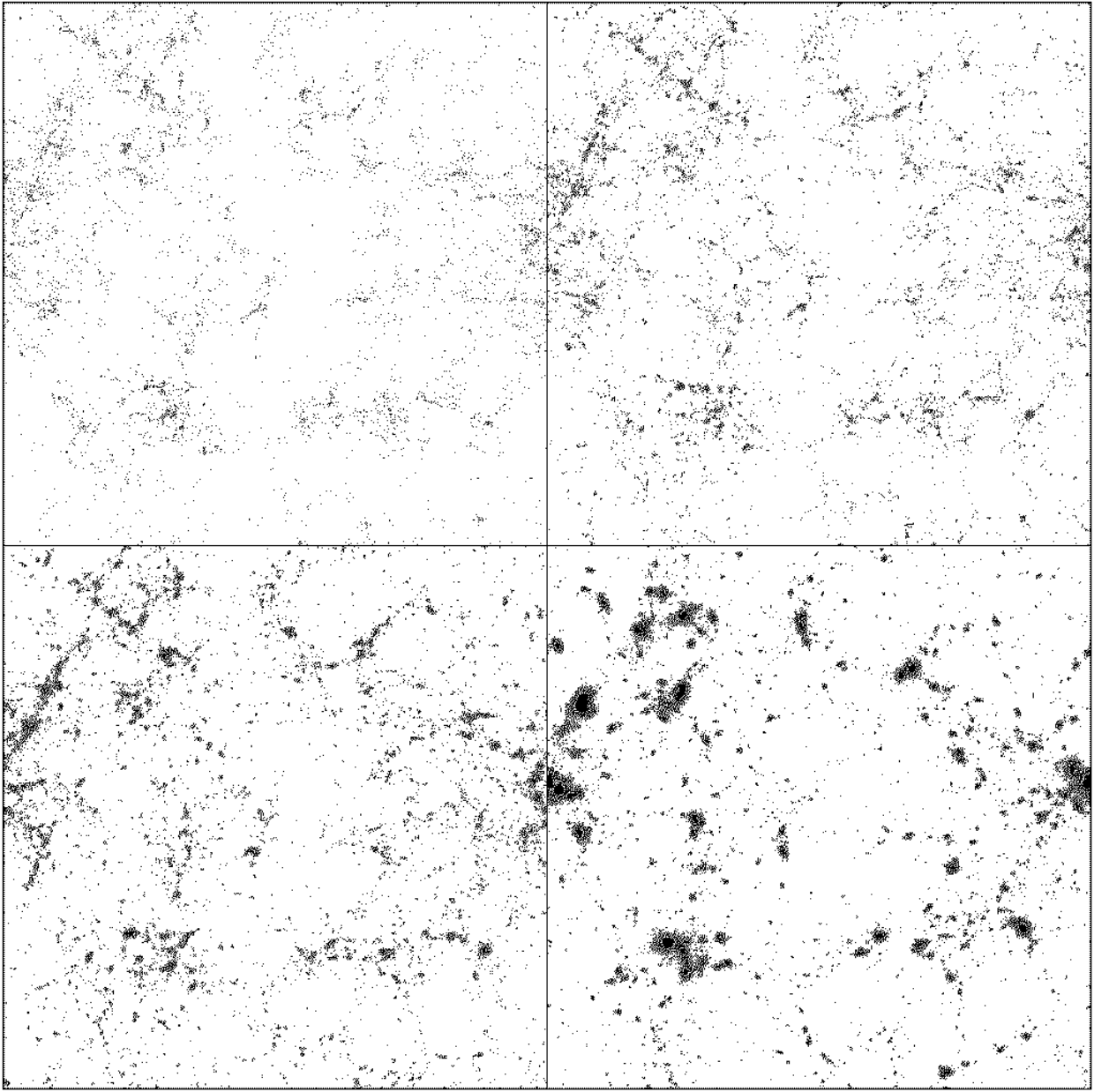}
\caption{The smoothed matter distribution at four redshifts, $z=5$
(upper left), 3 (upper right), 1.5 (lower left), and 0 (lower right),
from an $N$-body simulation of the $\onu=0.1$ C+HDM model.  Each panel
is 100$\times$100 Mpc comoving, and shows projection along one axis of
the entire simulation box. }
\end{figure}

%%%% Figure 3 %%%%
\begin{figure}
\epsfxsize=6.5truein 
\epsfbox{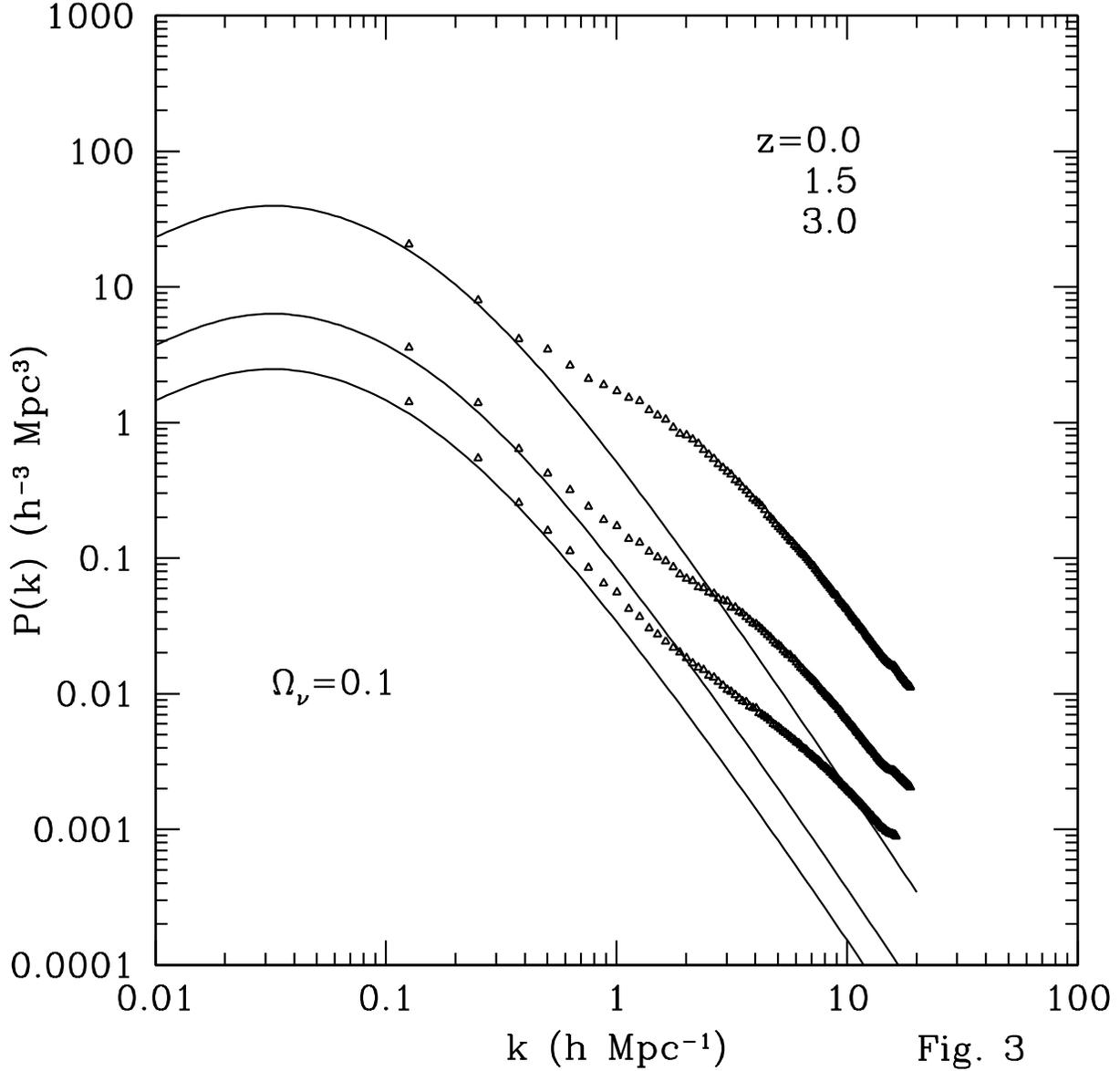}
\caption{The corresponding power spectrum at $z=3, 1.5$, and 0
(from bottom up), for the $\onu=0.1$ C+HDM model shown in Figure~2.  The
solid curves show the linear $P(k)$ predicted by the linear
perturbation theory; the triangles show the nonlinear $P(k)$ computed
from the $N$-body simulation.}
\end{figure}

%%%% Figure 4 %%%%
\begin{figure}
\epsfxsize=6.5truein 
\epsfbox{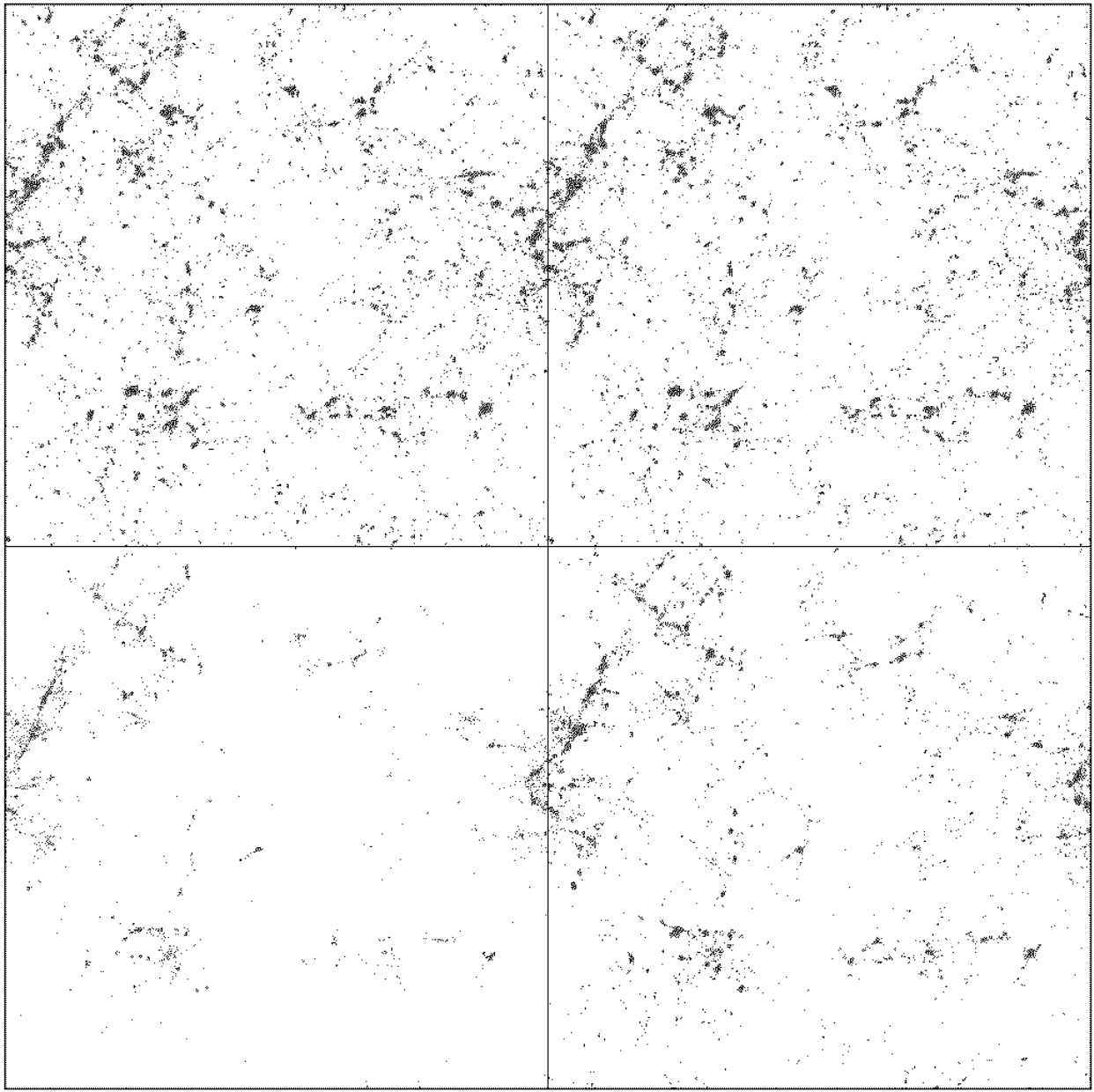}
\caption{The smoothed matter distribution at $z=1.5$ from $N$-body
simulations of four cosmological models: LCDM with $\om=0.3$ and
$\ov=0.7$, and C+HDM with $\onu=0.1, 0.2$, and 0.3 (clockwise from
upper left).  Each panel is 100$\times$100 Mpc comoving, and shows
projection along one axis of the entire simulation box.  The same
initial random phases were used in all four simulations, so structures
appear in similar locations in all panels.}
\end{figure}

%%%% Figure 5 %%%%
\begin{figure}
\epsfxsize=6.5truein 
\epsfbox{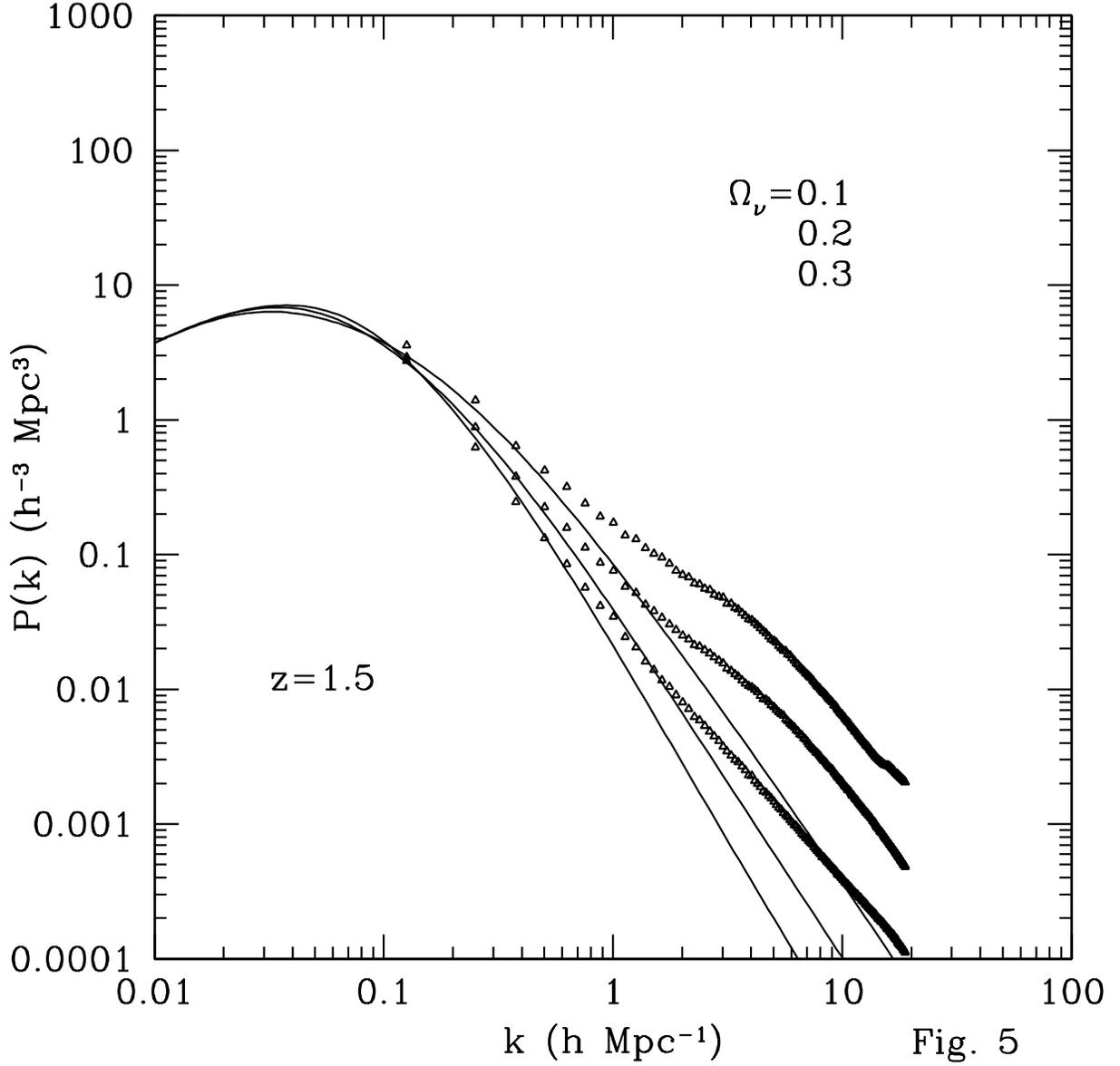}
\caption{The matter power spectrum at $z=1.5$ for the three C+HDM
models shown in Figure~4.  The solid curves show the linear $P(k)$
predicted by the linear perturbation theory; the triangles show the
nonlinear $P(k)$ computed from the $N$-body simulations.}
\end{figure}

%%%% Figure 6 %%%%
\begin{figure}
\epsfxsize=6.5truein 
\epsfbox{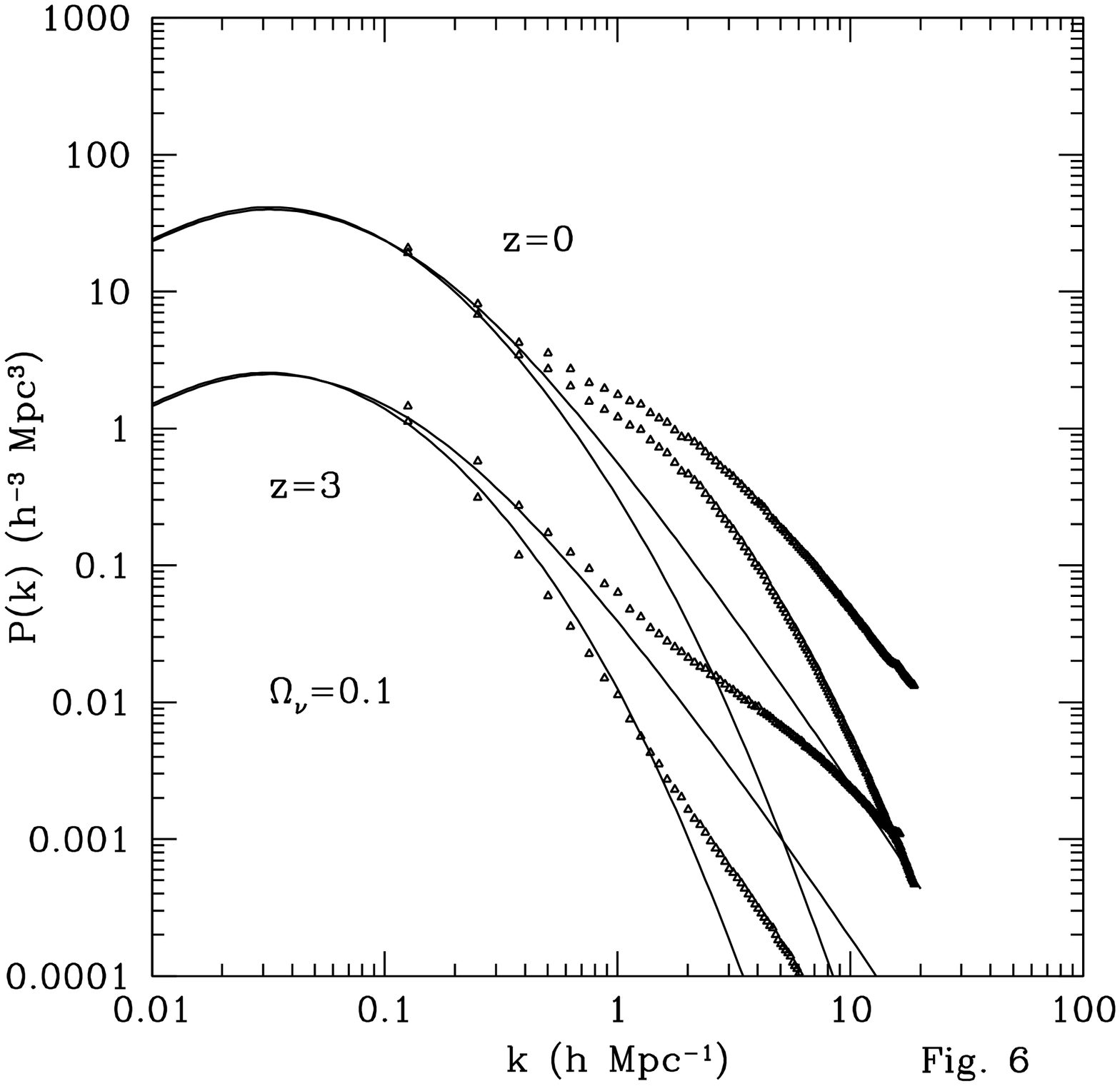}
\caption{The separate CDM and HDM power spectra at $z=3$ (lower
set) and 0 (upper set) for the $\onu=0.1$ C+HDM model.  The linear and
nonlinear $P(k)$ are shown as solid curves and triangles,
respectively.  For each pair of curves, the higher one is for the CDM
and the lower one for the HDM.}
\end{figure}

%%%% Figure 7 %%%%
\begin{figure}
\epsfxsize=6.5truein 
\epsfbox{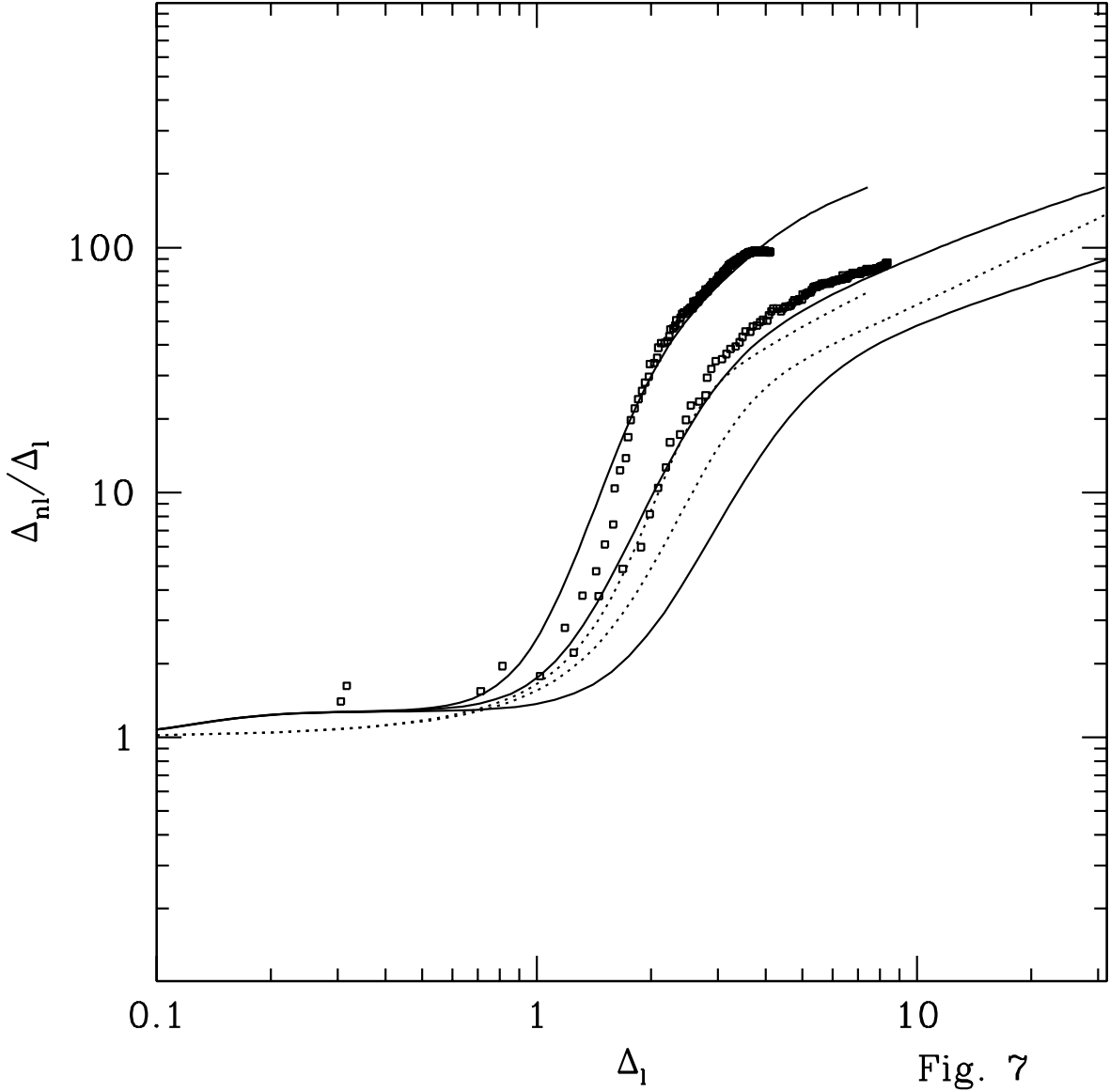}
\caption{The squares show the ratio of the nonlinear and linear
$\Delta(k)=4\pi k^3 P(k)$ at $z=0$ from the $\onu=0.2$ (left curve)
and 0.1 (right curve) C+HDM simulations.  The formula proposed by
Peacock \& Dodds (1996) for the pure CDM models leads to large errors
when applied to the C+HDM models (dotted curves; left for $\onu=0.2$
and right for 0.1).  The solid curves show our improved analytical
approximation from equation~(10), where the right-most one is for the
pure CDM model. (A modified approximation with a higher accuracy
than eq.~(10) has since appeared in Ma (1998).)}
\end{figure}

\end{document}